\documentclass[]{aastex631}
\usepackage{amsmath}

\begin{document}

\title{Laboratory investigation of the jet deflection: collisionless to strong-collision}

\correspondingauthor{Bin Qiao}
\email{bqiao@pku.edu.cn}

\author[0000-0002-9135-7173]{Zhu Lei}
\affiliation{Center for Applied Physics and Technology, HEDPS, and State Key Laboratory of Nuclear Physics and Technology,\\
School of Physics, Peking University, Beijing 100871, People’s Republic of China;}
\affiliation{Institute of Applied Physics and Computational Mathematics, Beijing 100094, People’s Republic of China;}

\author{Z. H. Zhao}
\affiliation{Center for Applied Physics and Technology, HEDPS, and State Key Laboratory of Nuclear Physics and Technology,\\
School of Physics, Peking University, Beijing 100871, People’s Republic of China;}

\author{X. Y. Li}
\affiliation{Department of Plasma Physics and Fusion Engineering, CAS Key Laboratory of Geospace Environment, University of Science and Technology of China, Hefei, Anhui 230026, People’s Republic of China;}

\author{L. X. Li}
\affiliation{Center for Applied Physics and Technology, HEDPS, and State Key Laboratory of Nuclear Physics and Technology,\\
School of Physics, Peking University, Beijing 100871, People’s Republic of China;}

\author{Y. Xie}
\affiliation{Center for Applied Physics and Technology, HEDPS, and State Key Laboratory of Nuclear Physics and Technology,\\
School of Physics, Peking University, Beijing 100871, People’s Republic of China;}

\author{W. Q. Yuan}
\affiliation{Center for Applied Physics and Technology, HEDPS, and State Key Laboratory of Nuclear Physics and Technology,\\
School of Physics, Peking University, Beijing 100871, People’s Republic of China;}

\author{B.Q. Zhu}
\affiliation{National Laboratory on High Power Laser and Physics, Chinese Academy of Sciences, Shanghai 201800, People’s Republic of China}

\author{J.Q. Zhu}
\affiliation{National Laboratory on High Power Laser and Physics, Chinese Academy of Sciences, Shanghai 201800, People’s Republic of China}

\author{S.P. Zhu}
\affiliation{Institute of Applied Physics and Computational Mathematics, Beijing 100094, People’s Republic of China;}

\author{X.T. He}
\affiliation{Institute of Applied Physics and Computational Mathematics, Beijing 100094, People’s Republic of China;}

\author{B. Qiao$^{*}$}
\affiliation{Center for Applied Physics and Technology, HEDPS, and State Key Laboratory of Nuclear Physics and Technology,\\
School of Physics, Peking University, Beijing 100871, People’s Republic of China;}

\begin{abstract}

The interaction between a jet and its surrounding medium plays a crucial role in various astrophysical phenomena, notably in star formation and jet deflection. Despite its significance, the detailed processes and fine shock structures involved in this interaction remain elusive due to limitations in current astronomical facilities, such as the confusing $\rm HCO^{+}$ emission in the Herbig-Haro (HH) 110/270 system. Here, we investigate the scaled plasma dynamics of this interaction under different collision states using laser-driven experiments. The experimental results demonstrate the importance of collision in jet deflection, revealing a distinctive double-shock structure in collision cases and a double-emission-shell structure in cases of strong collisions. Both cases exhibit a notable deflection of plasma jets by $50^{\circ}$ as a consequence of interaction with a high-density background. Optical images confirm that the impact point can be pushed into the inside part of the background medium, consistent with previous astrophysical models and simulations. In contrast, collisionless cases exhibit filament structures (the ion skin depth scale), yet there is an absence of discernible deflection in the direction of jet transmission. Furthermore, our findings have implications for understanding the shock structure in the HH 110/270 system. The significant $\rm HCO^{+}$ emission observed west of HH 110 may be attributed to the impact point being pushed into the inside part of the background.

\end{abstract}

\keywords{Laboratory astrophysics (2004), Plasma physics (2089), Young stellar objects (1834), Herbig-Haro objects (722), Interstellar medium (847), Stellar jets(1607)}


\section{Introduction} \label{sec:intro}

Supersonic collimated jets are widely observed in astrophysics \citep{reipurth2001, bally_protostellar_2016}. These jets are associated with various astrophysical objects, ranging from young stellar objects (YSOs) to active galactic nuclei (AGN), and are typically linked to driven source stars. However, there exist certain special jets, particularly some YSO jets (also known as Herbig-Haro (HH) objects), for which the corresponding driving source cannot be identified \citep{bally_protostellar_2016, Walawender2005}. Notably, HH 110 is an example of such a jet \citep{reipurth1986}. Despite extensive searches at optical, infrared, and sub-mm wavelengths, researchers have been unable to locate its driving source. However, they have discovered another jet, HH 270, which points towards the origin of the HH 110 jet \citep{reipurth1996}. The proper motion velocity ratio of HH 270 to HH 110 is approximately $2:1$, consistent with a deflection angle of approximately $60^\circ$. Consequently, they \citep{reipurth1996} proposed that HH 110 was formed through the deflection of the HH 270 jet as a result of its interaction with the surrounding medium. The mechanism behind this deflection remains a longstanding astrophysical mystery, and the more recent paper still focuses on the new deflection jets observed by the James Webb Space Telescope \citep{reiter2022}. Studies of the deflection jet have ruled out explanations based on magnetic fields \citep{hurka1999} or photo-ablation of jet surfaces \citep{bally2001}. Instead, the most likely cause is believed to be the effect of collision and ram pressure due to a side-wind \citep{raga1995}. Furthermore, enhanced emission and the presence of dense clouds have been observed in the vicinity of the HH 110/270 system \citep{choi2001, kajdivc2012}, providing additional support for the collisional explanation.

However, the physical process of the collision and the mechanism behind the generation of the deflected collimated jet is still a subject to debate. Theoretical analysis and numerical simulations suggest that an interaction between the jet and cloud should result in the formation of a double-layer shock structure \citep{yooung1991, raga1995}. The first shock deflects the incident jet in a new direction, while the second shock propagates into the background, compressing and heating the background cloud. However, some observational findings \citep{choi2001} contradict this jet-cloud collision model. These results indicate that only the deflected jet shows evidence of shock, with no detected shock emission in the impact region where the axes of HH 270 and HH 110 intersect. In contrast, a three-dimensional (3D) simulation \citep{raga2002} reveals the formation of a tunnel structure within the dense clouds due to the incident jet. The impact region is compressed towards the interior of the dense cloud by the backward shock in the background medium. This suggests that the impact region should be located further along the axis of the incident jet, which aligns with the significant $\rm HCO^{+}$ emission observed west of HH 110 \citep{choi2001}. However, further observational data is required to confirm this inference.

In addition to jet deflection, the interaction between the jet and the surrounding medium also plays a crucial role in various other phenomena, including star formation and the generation of cosmic rays. This interaction can initiate shock structures within the background, which in turn compress and heat the clouds. If the initial cloud is sufficiently dense, the positive feedback from these shocks can induce star formation \citep{mirabel2014, fragile2017, mukherjee2018}. In the case of ultrahigh-speed plasma jets, where the mean free path (MFP) may be much larger than the system size, the kinetic energy of plasma jets can only be dissipated through wave-particle collective interaction, resulting in the formation of a collisionless shock wave. This collisionless shock has the ability to accelerate charged particles to extremely high energies through the Fermi acceleration mechanism \citep{aharonian2004, fiuza2020}. Despite the wealth of information provided by the aforementioned observations and simulations regarding the interaction between the jet and medium, the fine shock structure and detailed interaction processes remain unclear due to the cross-scale features and limitations of observation facilities.

Over the past two decades, laboratory astrophysics has become a crucial tool in astrophysical research, thanks to advancements in high-power pulse technology \citep{remington_modeling_1999, remington2006, lebedev_exploring_2019}. It enables us to investigate astrophysical phenomena and validate related physical models. Despite the significant scale differences, the laboratory plasma can be scaled up to astrophysical objects using magneto-hydrodynamics scaling laws \citep{ryutov1999, ryutov2001}. Laboratory experiments have been conducted to study the interaction between the jet and medium, particularly regarding jet deflection. Lebedev et al. \citep{lebedev2004} examined plasma jet deflection using a pulsed power device and found that collimated jets can be deflected by the ram pressure of side winds. They observed that the deflection angle increased with enhanced ram pressure. Hartigan et al. \citep{hartigan2009} conducted jet deflection experiments at the Omega laser facility by interacting a laser-driven plasma jet with a spherical, low-density hydrocarbon obstacle. The experimental results revealed a chaotic structure of the deflected jet and the mixing of materials from the spherical obstacle. Yuan et al. \citep{yuan2015} demonstrated that a laser-driven plasma jet can be deflected by the ram pressure of a side wind, resulting in a deflection angle of $55^{\circ}$ when colliding with a nearby side wind. These experiments confirmed that collimated jets can indeed be deflected by the ram pressure of side winds, consistent with theoretical models \citep{raga1995}. However, these experimental results do not provide key evidence for the jet-cloud collision model, such as the presence of a double-layer shock wave structure and the inward shift of the impact region predicted in 3D astrophysical-scale simulations \citep{raga2002}.

In this paper, we present a comprehensive analysis of the interaction between a copper (Cu) plasma jet and a plastic (CH) background. Our analysis covers the entire process, ranging from collisionless to strong collision scenarios. In collisionless situations, filament structures emerge as a result of the Weibel instability. These filaments have an average width that closely aligns with the ion skin depth. However, they do not significantly alter the transport direction of the plasma jet. On the other hand, in collisional situations, the plasma jet experiences deflection due to the high-density laser-driven CH plasma core. The deflection angle is approximately 50 degrees. Additionally, we observe a bow-shaped working surface at the front of the deflected jet. Optical images clearly depict a double-shock structure, with the impact point being pushed inward into the CH background. These observations are consistent with astrophysical simulations conducted by Raga et al. \citep{raga2002}. Furthermore, the optical images also reveal a double-shock structure in the case of strong collisions. This structure is accompanied by a two-layer emission pattern that is evident in self-emitting X-ray data. These findings have been validated through radiation magneto-hydrodynamic (RMHD) simulations. Moreover, we find that the interaction cases exhibit electron acceleration to higher energy levels compared to the jet-only scenario. Considering the fulfillment of localization criteria and adherence to scaling laws, these experimental outcomes can be effectively utilized for the HH 110/270 system. These discoveries affirm the significance of collisions in deflecting astrophysical jets and demonstrate the possibility of shifting the impact point deeper into the background medium.

\section{Experimental setup}

\begin{figure*}
    \centering
    \includegraphics[width=16cm]{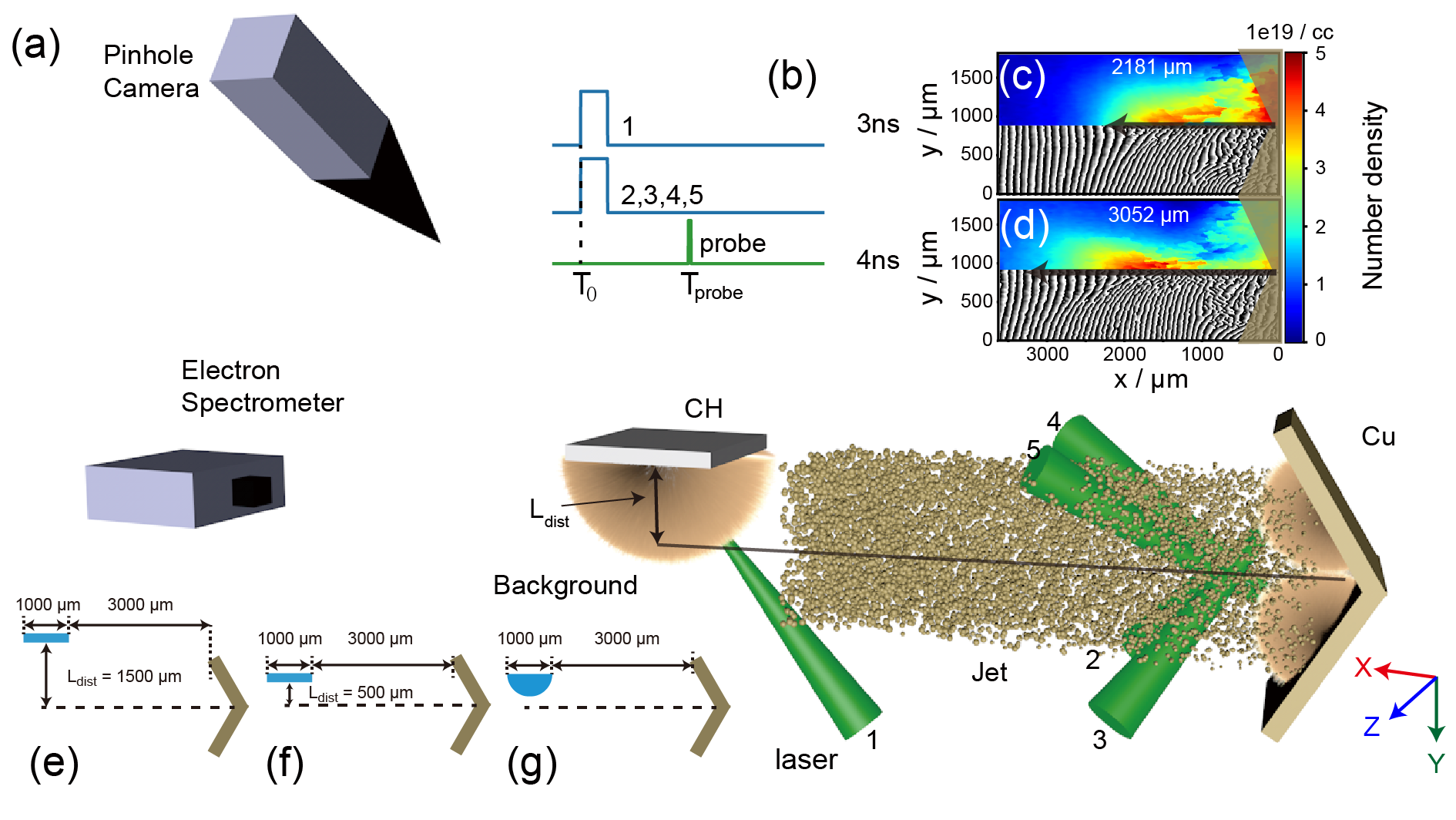} 
    \caption{{\bf Schematic view of the experimental setup.} Panel (a) illustrates the generation of the background plasma through the interaction between the laser and a hydrocarbon (CH) target (grey). The collimation jet is driven by nanosecond (ns) laser pulses (`2' - `5') from the V-shaped copper (Cu) target (golden). The distance between the CH target and the axis of the Cu target is denoted as $\rm L_{dist}$. The distance between the CH plane target and the Cu target is $3000 ~ \mu m$. The electron spectrometer is positioned in front of the Cu target. The pinhole camera is used to detect the X-ray radiation of the plasma. The evolution of the plasma is diagnosed using shadowgraphy and interferometry with a short probe beam. Panel (b) provides the time delays of different laser pulses, as well as the probe laser, for the optical Nomarski interferometer diagnostics. Panels (c) and (d) display maps of the inferred electron number density, in [$cm^{-3}$], at 3 ns and 4 ns respectively, obtained by inverting Nomarski interferometer images. The bottom parts of both panels show the corresponding interference fringes. Panels (e) to (g) illustrate the three target configurations used in our experiment, corresponding to collisionless, collision, and strong collision types respectively.}
    \label{expsetup}
\end{figure*}

The experiments were conducted at the ShenGuang-II (SG-II) laser facility in Shanghai, China. The experimental setup is illustrated in Fig.~\ref{expsetup}(a). The background medium is produced through the interaction between the laser and a hydrocarbon (CH) target, resulting in the emergence of a semi-spherical plasma outflow at the bottom of the CH target. The V-shaped targets consist of two $\rm 500 ~ \mu m$ thick Cu foils measuring $\rm 1 ~mm \times 1 ~ mm$, with an opening angle of $120 ^{\circ}$. Each foil is driven by two laser beams, generating a plasma plume on each foil. The convergence of these plasma plumes leads to the formation of a Cu plasma jet \citep{li2013}. The pinhole camera is used to detect the self-emitting X-rays of the plasma plumes, with the observation angle set at $\rm 45^\circ$ to the horizontal plane. The electron spectrometer is employed to record the electron energy spectrum. More details about the driven lasers and experimental targets are shown in Appendix \ref{app:expSetup}.

A probe laser is used to transversely pass through the interaction region (along the z-axis) to diagnose the plasma evolution using shadowgraphy and interferometry methods. The details of these methods are shown in Appendix Fig.~\ref{lasersetup}. The time delay between the probe laser and the driven laser is denoted as $T_{\text{probe}}$, as depicted in Fig.~\ref{expsetup}(b). The experiments were divided into two rounds, each with different probe beams. In the first round, the ninth beam of the SG-II laser facility was employed as the probe laser, with a wavelength of $\rm 527 ~nm$, a duration of 30 ps, and an energy of $\rm 15 \ J$. In the second round, the probe pulse had a duration of $\rm 6.5 \ ns$ and an energy of $\rm 15 \ mJ$. To match the diagnostic condition, the probe laser was shaped into a $\rm 120 \ ps$ pulse using a gated optical image (GOI). In the subsequent optical images, the gray images correspond to the first round, while the green images correspond to the second round. The number density distribution can be obtained by inverting the Nomarski interferometer images. As shown in the top parts of Fig.~\ref{expsetup}(c) and \ref{expsetup}(d), the number density of the Cu plasma jet is provided at 3 ns and 4 ns ($T_{\text{probe}} = T_0 + 3 \text{ or } 4 \ \rm ns$), and the inversion methods are referred to the paper \citep{park2016}. The corresponding interference fringes are displayed in the bottom parts of Fig.~\ref{expsetup}(c) and \ref{expsetup}(d). Based on the length of the jet at different probe times, the velocity of the Cu plasma jet is estimated to be approximately $\rm 800 ~ km/s$, with a mean number density of about $\rm 2 \times 10^{19} ~ cm^{-3}$.

Three different types of targets were utilized in our experiments, with the main distinction being the density of the CH background plasma during interaction with the Cu jet, resulting in different collision states. As depicted in Fig.~\ref{expsetup}(e) and (f), the distances between the CH plane target and the axis of the Cu target ($L_{\text{dist}}$) are $1500~ \rm \mu m$ and $500~ \rm \mu m$ respectively. According to the theory of steady-state planar ablative flow in a laser-produced plasma \citep{Manheimer1982}, the plasma density profile follows an exponential distribution. Consequently, the background number density in the two targets differs by approximately one order of magnitude. In Fig.~\ref{expsetup}(g), the plasma jet directly interacts with a high-density hemispherical CH target (without laser ablation), which has the smallest mean free path (MFP) among the three types of targets. Based on the results of numerical simulations and experiments, we refer to the three targets as the collisionless case, collision case, and strong-collision case respectively.

\section{Experimental results}

\subsection{CH Background Medium and Cu Jet.}

\begin{figure*}
    \centering
    \includegraphics[width=16cm]{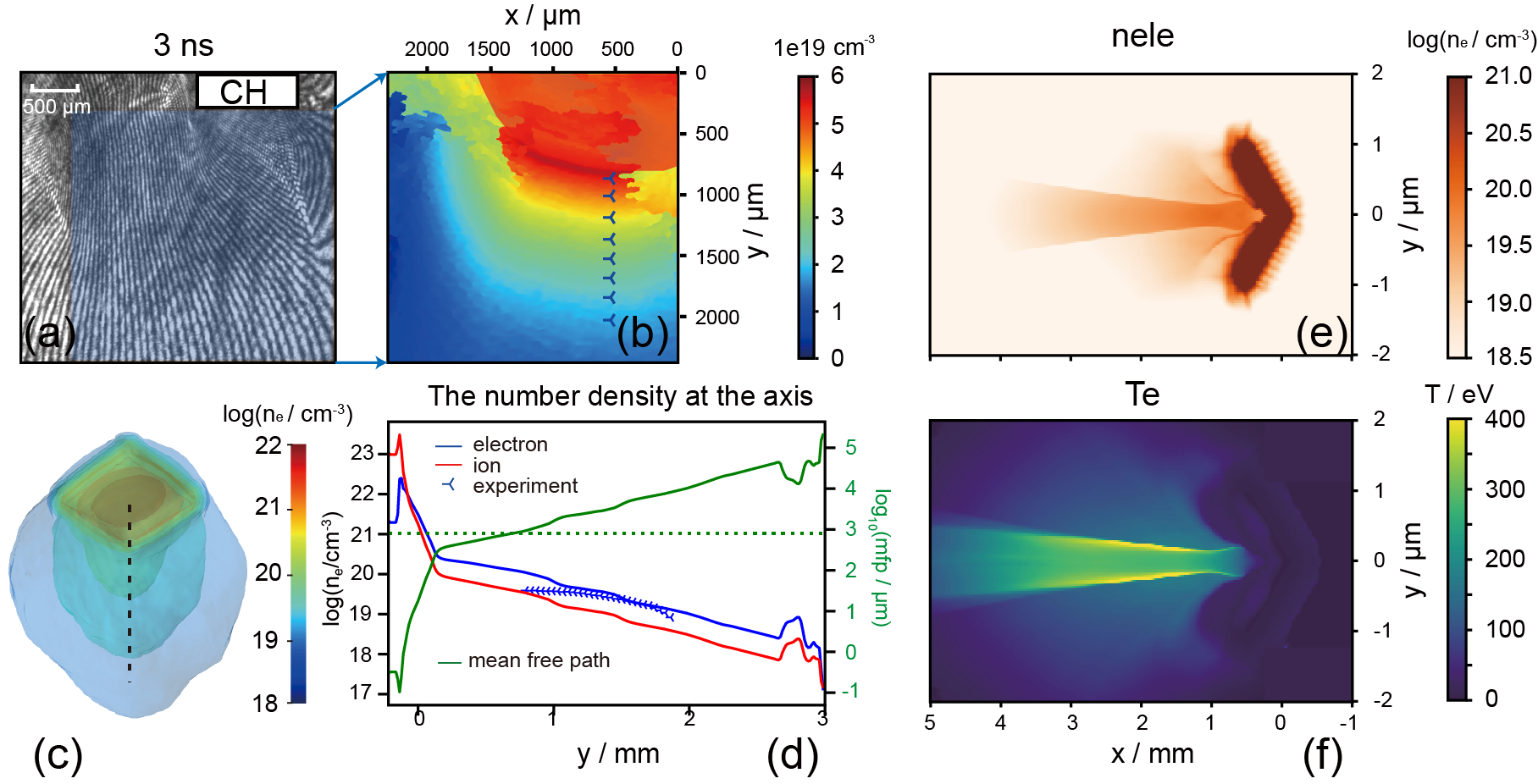} 
    \caption{\textbf{Free Expansion of CH Background Plasma and Cu Jet.} Panel (a) presents the interferometry image of the CH plasma at 3 ns, while panel (b) displays the corresponding number density map. To reproduce the expansion process, we utilized a three-dimensional radiation hydrodynamic simulation, as depicted in panel (c). The number density of electrons and ions in the axis, along with the mean free path of the Cu plasma jet, are plotted in panel (d). The experimental results for the number density in panel (b) are represented by triangle points, and the width of the jet ($800 \mu m$) is indicated by the green dot line. Panels (e) and (f) showcase the 2D slices of electron number density and temperature obtained from the 3D numerical simulation of the Cu plasma jet at 4 ns.}
    \label{back}
\end{figure*}

The evolution of the CH background plasma at 3 ns is illustrated in Figure \ref{back}(a), and the corresponding number density of the light blue region in panel (a) is presented in Figure \ref{back}(b). The CH plasma outflow exhibits nearly isotropic expansion, with the number density gradually decreasing radially outward from the center. Within a distance of approximately $500~\rm \mu m$ from the focal spot, the number density exceeds $1 \times 10^{20}~\rm cm^{-3}$. To reproduce the experimental findings, a three-dimensional radiation hydrodynamic simulation is conducted, and the simulation parameters are provided in Appendix \ref{app:simSetup}. As depicted in Figure \ref{back}(c), the lateral expansion of the plasma outflow is slightly slower than the vertical direction (perpendicular to the target surface). The number density at the axis (indicated by the black dot line in panel (c)) is presented in Figure \ref{back}(d). The triangle points represent the number density from the experiment in Figure \ref{back}(b). These results clearly demonstrate that the distribution of number density in the numerical simulation is roughly consistent with the experiment at the corresponding location. 

Regarding the Cu plasma jet, the 2D slices of electron number density and temperature are shown in Figures \ref{back}(e) and (f). The morphology of the jet closely resembles the experimental result, with a mean velocity of approximately $800~\rm km/s$. The mean temperature of the Cu jet is approximately $300~\rm eV$, corresponding to a sound speed $c_s$ of about $110~\rm km/s$ ($c_s = (\gamma Z k_B T/ m_i)^{\frac{1}{2}}$, where $\gamma = 5/3$ is the adiabatic index; $k_B$ represents Boltzmann’s constant; $Z\approx 15$, $T$, and $m_i = 64$ denote the mean charge state, temperature, and mass number of the jet, respectively). The Mach number $M_a = v / c_s$ is approximately $7$, indicating that the Cu plasma jet is supersonic.

In order to examine the impact of Coulomb collisions in the experimental setup, it is necessary to compare the mean free paths for collisions between the CH plasma background and the Cu plasma jet with the width of the jet \citep{lebedev2004,yuan2015}. The calculation details are provided in the section Appendix \ref{app:collParameter}. Figure \ref{back}(d) presents the plot of the mean free path, where the green dot line represents a mean free path of $800~\rm \mu m$, which is equivalent to the width of the jet. When the distance between the CH target and the axis of the Cu target, denoted as $\rm L_{dist} = 1500 \rm \mu m$ (labeled in Figure \ref{expsetup}(a)), the ion number density of the background medium is approximately $5 \times 10^{18} ~ \rm cm^{-3}$, resulting in a corresponding mean free path of about $6000~ \rm \mu m$. This value is significantly larger than the width of the plasma jet, indicating a fully collisionless interaction between the jet and the background. However, when $\rm L_{dist}$ is adjusted to $500 ~ \rm \mu m$, the ion number density reaches approximately $1 \times 10^{20} ~ \rm cm^{-3}$, and the mean free path reduces to about $315 ~\rm \mu m$, leading to a collisional interaction between the jet and the background. In the case of the jet interacting directly with a hemispherical CH target (Figure \ref{expsetup}(g)), the ion number density exceeds $1 \times 10^{22} ~ \rm cm^{-3}$, resulting in a mean free path of less than $5~\rm \mu m$. This indicates a strongly collisional interaction. In the experiments, the trigger time of the probe laser can be adjusted to observe the evolution of different collision states.

\begin{figure*}
    \centering
    \includegraphics[width=16cm]{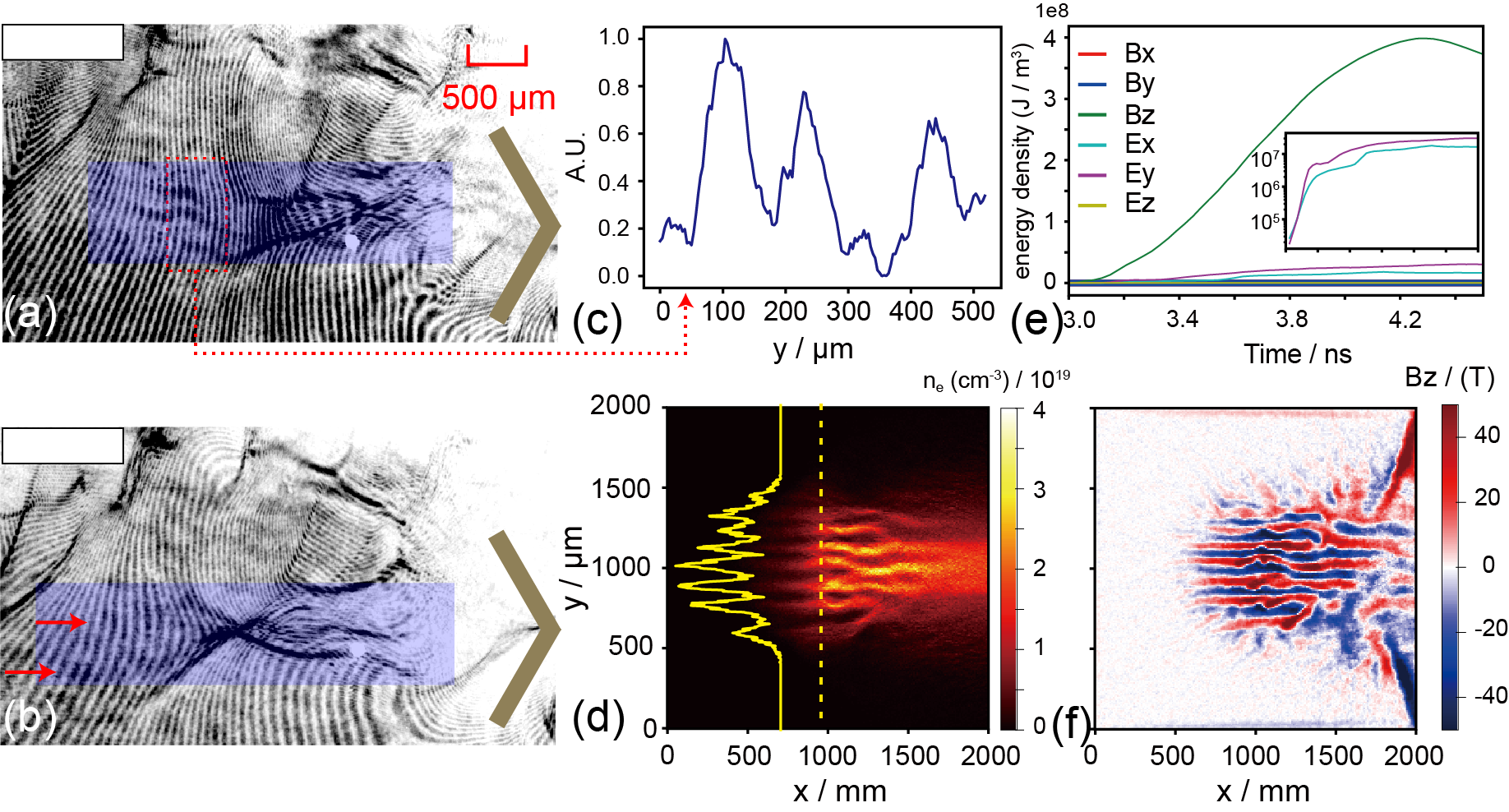} 
    \caption{\textbf{Collisionless Case and Simulations.} The interferometry images for the collisionless case are presented in panels (a) and (b), with the probe times set at $T_0 + 4~\rm ns$ and $T_0 + 5~\rm ns$, respectively. Panel (c) displays the average pixel data along the vertical direction of the red box in panel (a). In order to simulate the dynamics of plasma kinetic processes within the background, the particle-in-cell (PIC) method is employed for the collisionless scenario. The Cu plasma jet is injected from the left boundary of the box, while a uniform CH background is established within the box. The ion number density of the Cu plasma at $T_0 + 4~\rm ns$ is presented in panel (d), with the number density distribution along the y direction at the front of the Cu plasma jet indicated by the yellow line. The box-averaged energy densities of the electric field ($E_x,E_y,E_z$) and magnetic field ($B_x,B_y,B_z$) are displayed in panel (e), with the energy densities of $E_x$ and $E_y$ shown in the inset. The z-component magnetic fields (Bz) are plotted in panel (f).}
    \label{collisionless-case}
\end{figure*}

\subsection{Collisionless Case}
As discussed earlier, the collision state varies among the three types of targets due to the distribution of the CH plasma background. In this section, our focus is on analyzing the deflection of the Cu plasma jet and the physical process of energy conversion during the interaction between the jet and the background medium under different collision states. Figures \ref{collisionless-case}(a) and \ref{collisionless-case}(b) depict the collisionless case, where the distances between the CH plane target and the axis of the Cu target ($L_{dist}$) are set at $1500~\rm \mu m$. Due to the high asymmetry of the images, inverting the images to obtain effective density information is challenging. Therefore, we directly analyze the original data. In Figures \ref{collisionless-case}(a) and \ref{collisionless-case}(b), the yellow V-shaped structure and white rectangle represent the original positions of the Cu target and CH target, respectively, while the light-blue rectangles indicate the main regions of the Cu plasma jet. By comparing the images with different time delays, it can be observed that the Cu plasma jet is primarily transmitted along the axis of the V-shaped target without any deflection, even at $T_0 + 5~\rm ns$. In Figure \ref{collisionless-case}(a), filament-like structures appear at the front of the jet, with a mean full width at half maximum (FWHM) of approximately $100~\rm \mu m$, as shown in Figure \ref{collisionless-case}(c). The red arrows in Figure \ref{collisionless-case}(b) indicate that the filament structure gradually thickens.

For the fully collisionless scenario, the particle-in-cell (PIC) method is employed to simulate the dynamics of plasma kinetic processes within the background. To reproduce the experimental results in the collisionless case, we used typical data as the initial conditions for the PIC simulation. More details about this simulation can be found in Appendix \ref{app:PICSetup}. In the simulation, the Cu plasma jet is injected from the right boundary of the simulation box, with an electron number density of $\rm 2\times10^{19} ~ cm^{-3}$ and an average ionization of $Z_{Cu} = 15$. The simulation box contains a uniform CH background, with an electron number density of $\rm 5 \times 10^{18} ~ cm^{-3}$ and an average ionization of $Z_{CH} = 3.5$.

To improve the efficiency of the PIC simulation, a self-similar transformation is applied, as described in the Appendix \ref{app:PICSetup}. Figure \ref{collisionless-case}(d) illustrates the injection of a Cu plasma jet from the right boundary of the simulation box at 3 ns, corresponding to the time when the Cu plasma jet begins to interact with the CH background plasma. Panels (a) and (b) present the ion number density of the Cu plasma at 4 ns. According to the theory of collisionless plasma \citep{}, the appearance of filament-like structures is a distinct characteristic of Weibel instability development. In our experiments, the typical growth time of the Weibel instability is estimated as $\tau_w \approx c / (v_{flow} \omega_{pi}) = 0.08~\rm ns$. The simulation results demonstrate rapid filamentation at the front of the Cu plasma jet after its injection, as depicted in Figure \ref{collisionless-case}(d). The filaments exhibit a full width at half maximum (FWHM) of approximately $80~\rm \mu m$ (indicated by the yellow line in Figure \ref{collisionless-case}(d)), which is roughly equivalent to the ion skin depth and in good agreement with the experimental findings.

The box-averaged energy densities of the electric field and magnetic field are calculated using the following expressions \citep{stockem2009}: $\epsilon_{E_i} (t) = (N_x N_y)^{-1} \sum_{i, j} \epsilon_{0} [E_{i} (i \Delta_{x}, j \Delta_{y}, t)]^{2} / 2 $ and $\epsilon_{B_i} (t) = (N_x N_y)^{-1} \sum_{i, j} (B_{i} (i \Delta_{x}, j \Delta_{y}, t))^{2} / (2 \mu_0)$, where $E_i$ and $B_i$ represent different components of the electric and magnetic fields, respectively. The indices $i$ and $j$ correspond to the grid numbers in the x and y directions, and $\Delta_{x}$ and $\Delta_{y}$ represent the grid widths. $\epsilon_{0}$ and $\mu_0$ are the vacuum permittivity and vacuum magnetic permeability, respectively. Figure \ref{collisionless-case}(e) illustrates that the box-averaged energy density of $B_z$ initially increases exponentially and is significantly larger than the energy densities of other components. The inside figure of Figure \ref{collisionless-case}(e) shows a slight amplification of the energy densities $E_x$ and $E_y$. The energy densities of the other field components exhibit little growth. Furthermore, Figure \ref{collisionless-case}(f) demonstrates the appearance of positive and negative alternating filamentary structures of $B_z$ around the density filaments, which is a typical characteristic of Weibel instability.

In collisionless scenarios, the plasma jet propagates along the axis of the V-shaped target without any deflection, despite the ram pressure of the side winds being comparable to that of the jet. Through the mechanisms of Weibel instability and collective plasma effects, the plasma's kinetic energy is converted to field energy and internal energy. However, the efficiency of this energy conversion is relatively low, accounting for only about $0.6\%$ of the kinetic energy. Therefore, in collisionless cases, the influence of the background plasma on the collimated transmission of the jet is minimal.

\begin{figure*}
    \centering
    \includegraphics[width=16cm]{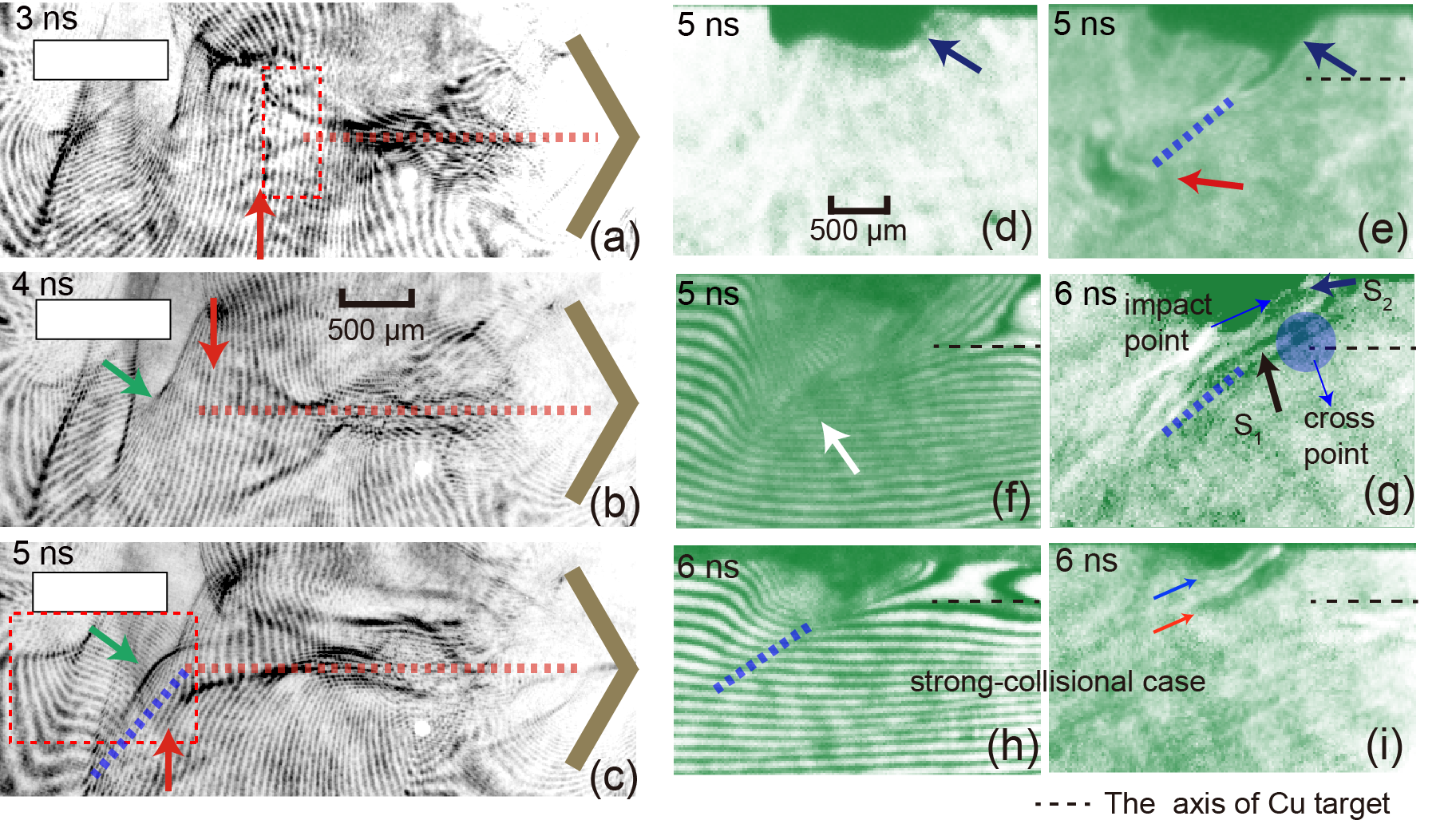} 
    \caption{\textbf{Collisional and strong-collisional experimental results.} The presented experimental results are divided into two rounds. Panels (a) to (c) display the experimental Nomarski images from the first round, obtained using the ninth beam of the SG-II laser facility, with probe times of $T_0 + 3,4,5 ~ \rm ns$, respectively. Panels (d) to (i) show the experimental results from the second round, focusing on the evolution of the deflected jet and the CH background. Panel (d) illustrates the shadowgraphy of the pure CH background at $T_0 + 5 ~ \rm ns$, revealing a high-density core at the bottom of the CH target. Panels (e) and (f) present the shadowgraphy and Nomarski images, respectively, for the collisional case at $T_0 + 5 ~ \rm ns$. The probe time for panel (g) is $T_0 + 6 ~ \rm ns$. Panels (h) and (i) display the experimental images for the strong-collisional case at $T_0 + 6 ~ \rm ns$. In panels (d) to (i), the black dotted lines indicate the position of the axis of the Cu target, while the blue dot lines represent the axis of the deflected jets.}
    \label{collision-case}
\end{figure*}

\subsection{Collisional and Strong-collisional Cases}

When the distance between the CH plane target and the axis of the Cu target is adjusted to $500~ \rm \mu m$, the interaction between the plasma jet and the background medium becomes collisional. Fig.~\ref{collision-case}(a) to \ref{collision-case}(c) provide a direct observation of the global dynamic evolution of this interaction. The plasma density profile of the CH background plasma follows an exponential distribution, indicating a transition from collisionless to collision in the interaction between the Cu plasma jet and the background medium. At $T_0 + 3$ ns (Fig.~\ref{collision-case}(a)), the Cu jet encounters the CH background at approximately $1800 ~\mu m$ from the center of the CH target, with a mean free path (MFP) of about $6000 ~ \mu m$, which is much larger than the jet width. Filament-like structures appear at the front of the jet, with a full width at half maximum (FWHM) of approximately $\rm 80 ~\mu m$, similar to the collisionless case. As the jet continues to penetrate the background, the number density of CH plasma gradually increases, and the MFP decreases to $\rm 1000 \sim 2000  ~ \mu m$, indicating a weak collision state. Fig.~\ref{collision-case}(b) confirms that the direction of jet propagation remains consistent without significant deviation. However, when the Cu plasma jet interacts with the high-density core, the dynamics of the jet differ significantly from the collisionless situation. A comparison between panels (b) and (c) reveals a substantial alteration in the shape of the high-density CH region indicated by the green arrow, along with compression of the high-density background and the appearance of a strong shock. The MFP in this region is approximately $\rm 100 \sim 300 \rm ~ \mu m$, representing a collisional state. Due to strong collision, the jet deflects downward at an angle of approximately $50^{\circ}$.

In the second round, our focus is primarily on the later stages of the deflected jets and the background medium. Fig.~\ref{collision-case}(d) shows the shadowgraphy of the free expansion of the CH background at $T_0 + 5 ~ \rm ns$. The shadowgraph systems are used to detect shock waves, which are characterized by large density gradients, by indicating the variation of the second derivatives of the index of refraction. When the plasma jet interacts with the background, the high-density core of the CH plasma is compressed, resulting in the appearance of a bow-shaped working surface in the direction of jet deflection (as indicated by the red arrow in Fig.~\ref{collision-case}(e)). The Nomarski interferometer image (Fig.~\ref{collision-case}(f)) reveals that the deflected jet maintains its collimation structure. At $T_0 + 6 ~ \rm ns$ (Fig.~\ref{collision-case}(h)), a double-shock structure is formed (indicated by the black and blue arrows), and the Cu plasma jet is deflected along the first shock wave. The high-density CH background is significantly compressed by the second shock wave, with a compression speed of approximately $\rm 150 ~ km/s$, as calculated by comparing the edge of the high-density region in Figures \ref{collision-case}(e) and \ref{collision-case}(f). The light blue region represents the axis of deflected jet that crosses the axis of the original incident jet, but the impact point does not lie within this region. Fig.~\ref{collision-case}(g) shows that the impact point is pushed inside the CH background. This experimental result is consistent with astrophysical-scale 3D simulation results \citep{raga2002}.

In the strong-collisional case, the plasma jet directly interacts with a hemispherical CH target, as shown in Fig.~\ref{collision-case}(h) and \ref{collision-case}(i). Since there is no laser irradiation, the spatial scale of the background medium is much smaller compared to the previous cases, resulting in only a portion of the plasma jet interacting with the high-density background. Interestingly, we still observe a double-shock structure in the shadowgraphy image (indicated by the blue and red arrows in Fig.~\ref{collision-case}(i)), similar to the collisional case. However, the difference lies in the fact that the CH background is hardly compressed by the Cu plasma jet. From the analysis of these experimental results, we can conclude that the dynamic evolution of the plasma jet is closely linked to the collision state, particularly in the collisionless and collision scenarios. Additionally, the compression velocity of the CH background is influenced by the background plasma density.

\begin{figure*}
    \centering
    \includegraphics[width=14cm]{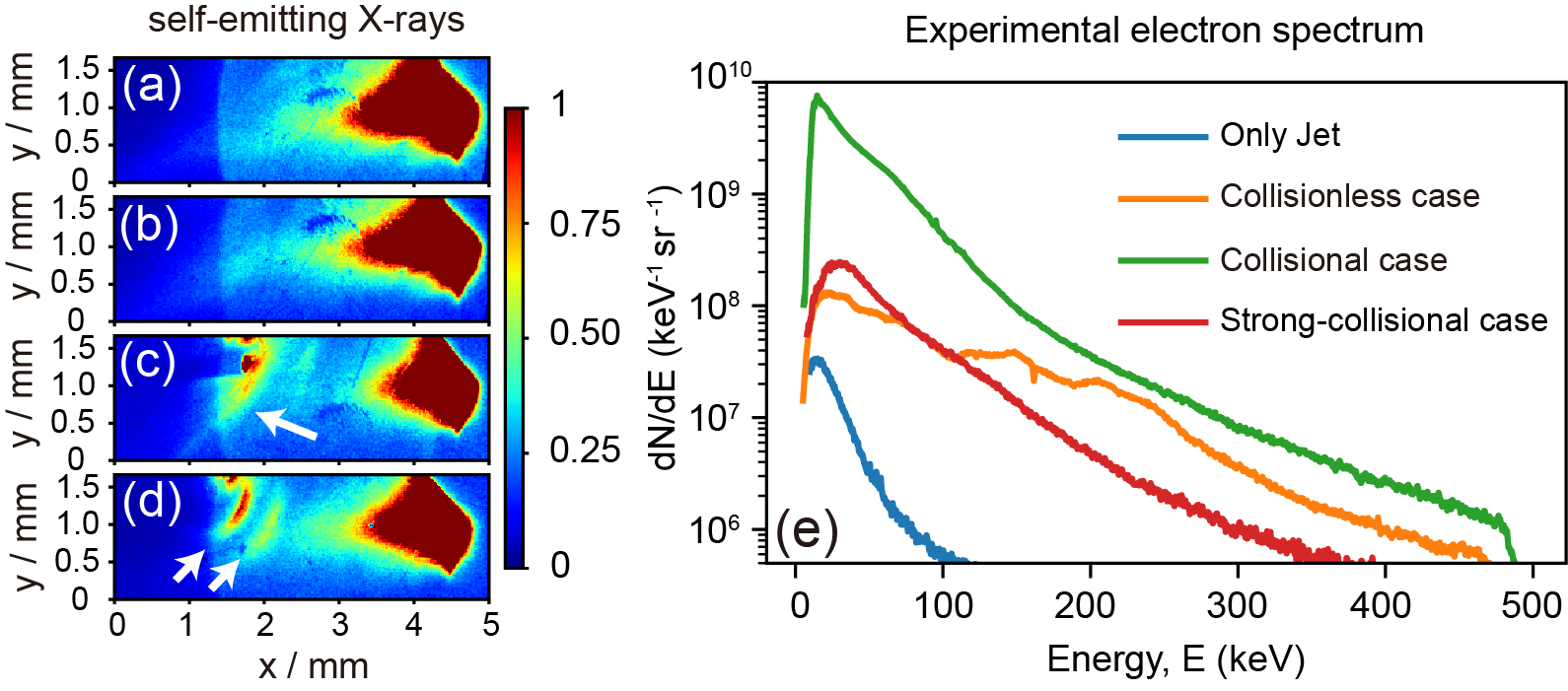} 
    \caption{{\bf Self-emitting X-rays and electron spectrum.} Panels (a) to (d) are the images of self-emitting X-rays through a pinhole camera. Panel (a) is the X-ray self-emission of the Cu plasma jet, and panel (b) corresponds to the collisionless case. Panel (c) and (d) are corresponding to the collisional and strong-collisional cases. The experimental time-integrated electron spectrum of different cases is given in panel (e).}
    \label{xray-spectrum}
\end{figure*}

\subsection{X-ray Self-emitting Images and Electron Spectrum}

The self-emitting X-rays for different cases are captured using the pinhole camera, as shown in Fig.~\ref{expsetup}(a). The magnification is set to $\times 6$, and a $\rm 2 ~ \mu m$-thick aluminum (Al) foil filter is utilized to allow X-rays with energies above $\sim 400 ~ \rm eV$ to pass through. The corresponding experimental results are presented in Fig.~\ref{xray-spectrum}(a) to \ref{xray-spectrum}(d). In the measurement region, the bulk plasma is assumed to follow a Maxwell-Boltzmann distribution, and the primary radiation mechanism for X-rays is bremsstrahlung. The bremsstrahlung emission power per unit energy and volume is scaled as $J_{br}(\nu, T) \propto T^{-\frac{1}{2}} n_e n_i Z^2 e^{-h\nu / kT} \ \rm (erg ~ s^{-1} ~ cm^{-3} ~ Hz^{-1})$ \citep{halverson1972}, where $\nu$ and $T$ represent the photon frequency and temperature, respectively. Additionally, $n_e$, $n_i$, and $Z$ denote the number density of electrons, ions, and the average ionic charge. By integrating the emission over the entire spectrum, we obtain the power per volume scaling as $T^{1/2} n_e n_i Z^2 \ \rm (erg ~ s^{-1} ~ cm^{-3})$ \citep{rybicki1991}. Hence, we can conclude that the luminous intensity is primarily influenced by the plasma temperature and density.

Fig.~\ref{xray-spectrum}(a) displays the X-ray self-emission of the Cu plasma jet, where a collimated emission structure is observed along the axis of the V-shaped Cu target, with a length of approximately $\rm 3000 ~ \mu m$. In the collisionless case (Fig.~\ref{xray-spectrum}(b)), the emission morphology closely resembles the result shown in Fig.~\ref{collisionless-case}(a), indicating minimal interaction between the plasma jet and the background. Fig.~\ref{xray-spectrum}(c) reveals a luminous bright region, approximately $600 ~ \mu m$ in length, beneath the CH target in the collisional case. This region is positioned close to the first shock ($S_1$ in Fig.~\ref{collision-case}(g)). The intensified X-ray self-emission beneath the CH plane target suggests the formation of a strongly compressed and heated zone in the interaction region. In the strong-collision case, where the jet interacts with a CH hemisphere target, Fig.~\ref{xray-spectrum}(d) exhibits a double-layer emission structure, consistent with the shadowgraph image at $T_0 + 6 ~ \rm ns$ (Fig.~\ref{collision-case}(i)). This curved double-layer structure provides further evidence for the formation of two stable shock zones in front of the CH hemisphere target.

The time-integrated electron spectrum for different cases is displayed in Fig. \ref{xray-spectrum}(e). A comparison between the measured spectrum in the jet-only case (blue line in \ref{xray-spectrum}(e)) and the interaction case reveals significant differences. In the jet-only case, the measured electron spectrum shows a much lower number ($< 20.6 \%$) and energy ($< 16 \% $) of electrons above 40 keV. And there is almost no signal for the background-only case. Conversely, in the interaction cases, the thermal temperature (peak position) of electrons is higher, and the electrons can be accelerated to higher energies. Therefore, our electron spectrometer measurements indicate the presence of acceleration mechanisms during the interaction. Previous experiments and simulations \citep{huntington2015,li2019,fiuza2020} have shown that charged particles can be accelerated by the Weibel-dominated electromagnetic shock in collisionless plasma. Our experiment results also reveal the presence of filament structures, suggesting that electrons may be accelerated by the electromagnetic structures induced by the Weibel instability. However, it is important to note that a collisionless shock is not formed in our experiments. In the case of collisions, electrons may be accelerated by vortex-like magnetic field cloud structures induced by the Kelvin-Helmholtz (KH) instability through the Fermi acceleration mechanism \citep{zhang2009}. More details on this are shown in the subsequent radiation magneto-hydro simulation results (see Fig. \ref{xray-spect}(a)). The lower cut-off energy and quantity in the strong collision case compared to the collision case may be due to the suppression of the KH instability. All the aforementioned inferences are based on our experimental data and simulation results, and we will continue to analyze the acceleration mechanisms in future research.

According to the aforementioned experimental results, the morphology and radiation characteristics of the interaction between the jet and background differ in the collision and collisionless scenarios. Specifically, in the case of collision and strong collision, collisions dominate the interaction zone. Therefore, we employ the radiation magneto-hydrodynamics method to replicate the experimental findings. Furthermore, additional insights are provided by the simulation results.

\section{RMHD simulations for collision situation}

In order to study the interaction between jets and background plasma, we conducted 3D numerical simulations using the radiation magneto-hydrodynamics code FLASH \citep{fryxell2000}. Further details can be found in Appendix \ref{app:simSetup}. Figures \ref{simu-hydro}(a) to (d) depict the entire process of interaction between the Cu jet and CH background, with a distance $\rm L_{dist}$ set at $\rm 500 ~ \mu m$. The color maps represent the logarithmic mass density at different times. The total pressure, denoted as $\rm P_{tot}$, is defined as the sum of the thermal pressure ($\rm P_{thermal}$) and the ram pressure ($\rm P_{ram}$), given by $P_{tot} = P_{thermal} + P_{ram} = nk_BT + \rho v^2$. Here, $n$ represents the number density, $k_B$ is the Boltzmann constant, $T$ is the temperature, $\rho$ is the density, and $v$ is the velocity. The total pressure of the Cu plasma jet and CH background at the axis of the V-shape target, indicated by the dashed line in Figure \ref{simu-hydro}(a) to \ref{simu-hydro}(c) (ranging from $\rm x = 3 ~ mm$ to $\rm x = 5 ~ mm$), are plotted by the black and white lines. In the FLASH code, the Cu and CH plasmas are differentiated by their respective species names. Additionally, a shock detection method, as described by Balsara et al. (1999) \citep{balsara1999}, is implemented in the FLASH code. If a strong shock structure is detected, the shock parameter is set to 1. 

\begin{figure*}
    \centering
    \includegraphics[width=17cm]{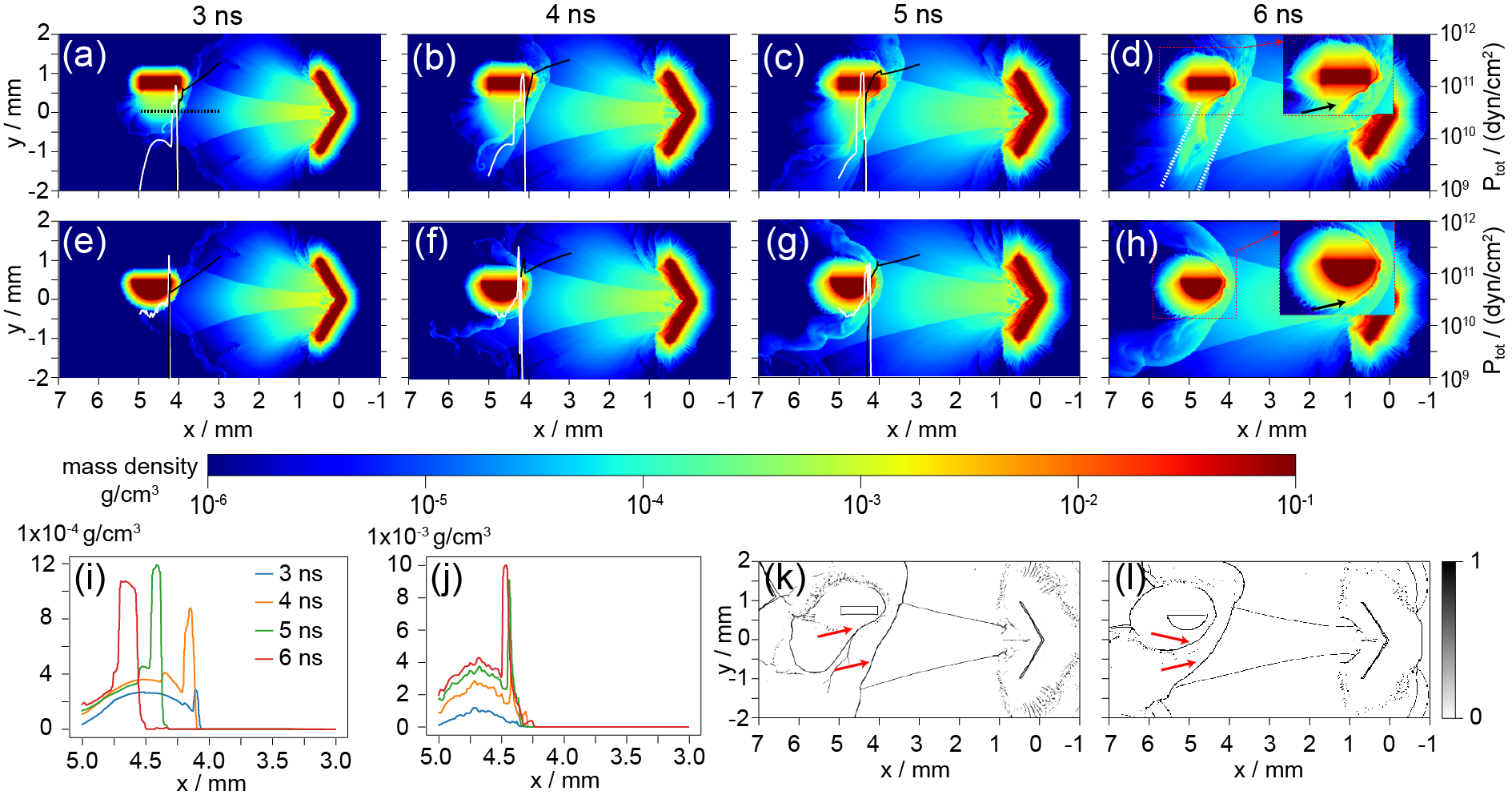} 
    \caption{{\bf The simulation results for the collision and strong collision case.} Panels (a) to (d) display the interaction results of the collision case, with a distance of $\rm L_{dist} = 500 \mu m$. Panels (e) to (h) illustrate the strong collision case. And the color maps depict the logarithmic density at different times. In panels (a) to (h), the black and white lines represent the total pressure of the Cu plasma jet and CH background, respectively, at the axis of the V-shape target ($\rm y = 0$) from $\rm x = 3 ~ mm$ to $\rm x = 5 ~ mm$ (indicated by the black dot line in panel (a)). The mass density of the CH background at the same position is shown in panels (i) (for collision) and (j) (for strong collision). Panels (k) and (l) indicate the positions where shocks were detected by the program for the two cases at 6 ns, respectively.}
    \label{simu-hydro}
\end{figure*}

At 3 ns, the low-density part of the jet has reached the edge of the high-density CH background in the collision case. Subsequently, the CH background undergoes compression due to the higher total pressure of the Cu plasma jet. The interaction region exhibits a double-shock structure based on the shock parameter analysis, as depicted in Figure \ref{simu-hydro}(k). The first shock arises from the collision between the supersonic plasma jet and the background, resulting in the formation of a mixed interaction region (indicated by white dot lines in Figure \ref{simu-hydro}(d)). Consequently, the plasma jet is deflected downward along the direction of the first shock. The pressure within the shocked CH plasma exceeds that of the undisturbed zone as a result of its higher density and temperature. Additionally, a high-density shell is formed at the contact surface. Simultaneously, the velocity of the shocked CH plasma decreases, causing the unshocked region to move faster and collide with the shell. This collision takes place at supersonic speeds, leading to the formation of a shock wave. Consequently, we observe the propagation of a backward shock that compresses the CH background inward. The inner figure of Figure \ref{simu-hydro}(d) depicts a backward shock that compresses the CH background medium. The mass density of the CH background increases by 2 to 3 times compared to the undisturbed zone when the forward shock passes, as illustrated in Figure \ref{simu-hydro}(i). Additionally, the peak mass density of CH gradually moves inward with an estimated velocity of $\rm 250 ~ km/s$. By considering the equilibrium between the post-shock jet and background plasma, the speed of the shock in the background can be estimated \citep{reipurth1996,raga1995} using the equation $v_{back} = (\rho_{jet} / \rho_{back}) ^{1/2} v_{jet}$, where $\rho_{jet}$ and $\rho_{back}$ represent the mass density of the Cu jet and background, respectively. According to the simulation, the mass density ratio of the jet to the background is approximately 0.12, yielding an estimated speed of $\rm 260 ~ km / s$, which is consistent with the experimental findings. 

In the case of strong collision, the solid CH target undergoes gradual ionization due to the thermal radiation generated by the high-temperature Cu plasma (Figure \ref{simu-hydro}(e)). Similar to the previous case, a strong shock is formed when the plasma jet interacts with the high-density background. However, in this case, the Cu plasma jet is unable to move inward due to the higher pressure of the CH plasma (Figure \ref{simu-hydro}(e) to (g)). At the first strong shock, the Cu plasma jet is deflected downward at an angle of approximately $\rm 45 ^{\circ}$, which is consistent with experimental observations. Additionally, a backward shock is also formed within the CH plasma, but its velocity is only about $\rm 30 ~ km / s$. In front of the solid hemisphere target, a nearly quasi-static high-density structure is formed, as indicated by the black arrows in Figure \ref{simu-hydro}(h). This observation aligns with both experimental findings (Figure \ref{collision-case}(i)).

\begin{figure*}
    \centering
    \includegraphics[width=8cm]{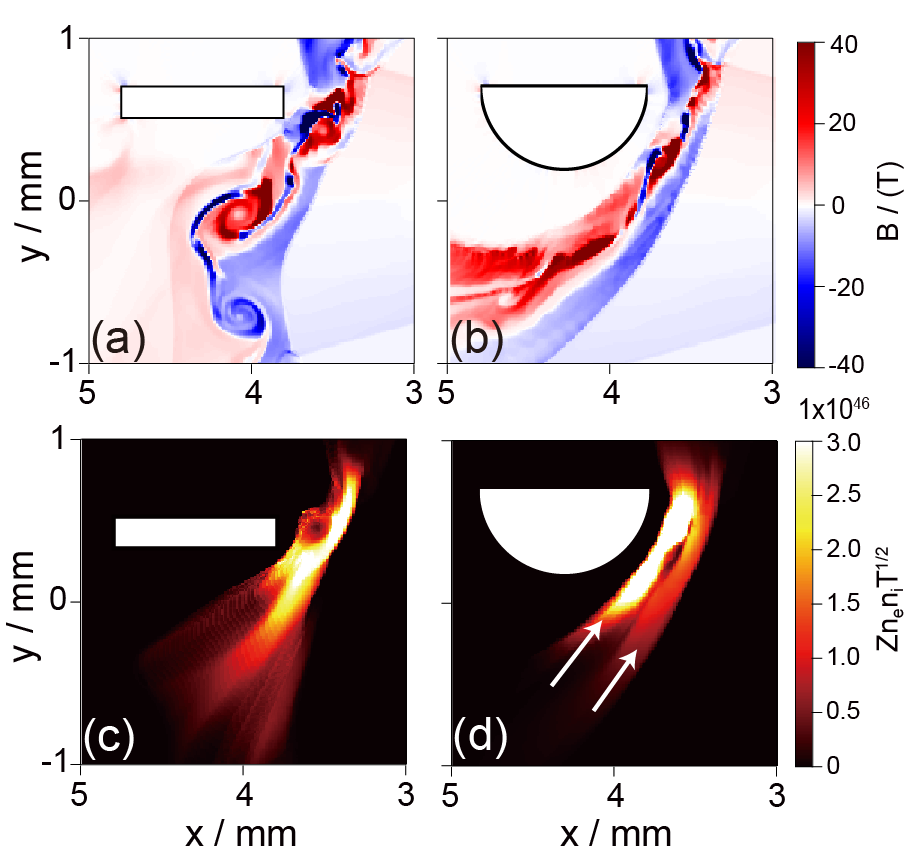} 
    \caption{{\bf Integrated X-ray emission and self-generated magnetic at the interaction zone} The self-generated Biermann magnetic fields at the interaction zone for collision and strong collision cases are shown in panels (a) and (b) respectively. We integrate the $T^{1/2} n_e n_i Z^2 $ at different moments to demonstrate the reason for the appearance of a double-shell structure in the self-emitting X-ray data in the experiment, as shown in panels (c) and (d) respectively. The white area represents the initial position of the target.}
    \label{xray-spect}
\end{figure*}

The self-generated Biermann magnetic fields at the interaction zone for both cases are illustrated in Figure \ref{xray-spect}(a) and \ref{xray-spect}(b). The plots clearly demonstrate the amplification of magnetic fields at the interface due to the compression induced by the plasma flow. Moreover, the KH instability experiences rapid growth as a consequence of the discrepancy in downward velocities at the interface of the CH background and Cu plasma jet. Figure \ref{xray-spect}(a) reveals the presence of a vortex structure at the interface, with significantly higher magnetic fields within the KH vortex compared to other regions, resulting from the vortex motion of the plasma. This observation suggests that the self-generated magnetic fields are amplified through the KH instability \citep{alves2012}, which may lead to the acceleration of charged particles \citep{borse2021}. In the case of strong collisions, the mass density of the CH plasma background is one order of magnitude higher compared to the previous case. As a result, the critical unstable mode of the KH instability, expressed as $[g\left(\rho_{2}^{2}-\rho_{1}^{2}\right)] / [\rho_{1} \rho_{2}\left(U_{2}-U_{1}\right)^{2}]$, indicates that the large-wave KH instability is suppressed due to the significantly higher density background. The analysis of the self-generated fields also reveals the absence of large-scale KH vortices.

After the passage of the shock wave, the temperature of the shocked region undergoes a rapid increase, leading to a corresponding increase in the self-emitting X-rays emitted by the plasma. As discussed earlier, the power per unit volume of self-emitting X-rays scales with $T^{1/2} n_e n_i Z^2 \ (erg ~ s^{-1} ~ cm^{-3})$. Therefore, by integrating the relevant data at different time points, we can obtain a similar structure compared with the self-emitting X-rays, which can further explain the formation of the double-shell structure. The detailed integration method is described in Appendix \ref{app:Xrays}. Figure \ref{xray-spect}(c) and (d) present the integrated data. In the case of collision, a strong emission layer is formed in front of the CH plain target, which coincides with the location of the first shock wave. However, the integrated emission of the backward shock within the CH background is not as strong due to the high compression velocity. On the other hand, in the case of strong collision, a double-layer emission structure is observed, with the inner layer exhibiting significantly higher intensity. This emission structure is consistent with the two stable shocks depicted in Figure \ref{simu-hydro}(l). The presence of the strong inner emission layer can be attributed to the gradual backward shock within the CH plasma. Importantly, both emission characteristics align with the experimental results shown in Figure \ref{xray-spectrum}(c) and \ref{xray-spectrum}(d).

\section{Discussion}

\begin{table*}
\setlength{\tabcolsep}{7mm}{    
      \caption{Key parameters of laboratory and HH 110/270 jets under scaling laws\label{scaleparameters}}

      \begin{tabular}{lccr}
      \hline
      Parameters                     & Laboratory jets           & Scaled-up jets                              & HH 110/270 jet       \\
      \hline
      
      Reynolds number                & $\sim 8\times10^5$                & $\cdots$                                   & $> 10^8$     \\
      Peclet number                  & $\sim 30$                   & $\cdots$                                   & $\sim 10^{7}$ \\
      Temperature                    & $\sim 300 \ \rm{eV}$   & $\cdots$                             & $\sim 1 - 10 \ \rm{eV}$ \\
      Length                         & $\rm \sim 0.3 \ cm$        & $\rm  \sim 10^{15} ~ cm$                   & $ \rm \sim 10^{15} \ cm $    \\
      Number density                 & $ 10^{19}\ - \  10^{20} \ \rm{cm}^{-3}$  & $\sim 2 \times (10^2-10^3) \ \rm{cm}^{-3}$   & $10^{1} \ - \ 10^{3} \ \rm{cm}^{-3}$   \\
       Pressure                       & $10^{8} \ - \ 10^{9} \ \rm{Pa}$   & $ 3\times (10^{-10} \ - \ 10^{-9}) \ \rm{Pa}$       & $10 ^{-11}\ - \ 10^{-9} \ \rm{Pa}$  \\
      Velocity                       & $\rm \sim 800 \ km/s$      & $\sim 300 \ \rm{km}/\rm{s}$          & $\rm 200-500 \ km/s$    \\
      Timescale                      & $\sim 5 - 10 \rm{ns}$      & $ \rm \sim 2 - 4 \ years $              & $ \rm 10 \ years $  \\
      
      \hline
      \end{tabular}\\
  } 
\end{table*}

The interaction between the jet and the surrounding medium plays a crucial role in various astrophysical phenomena. In this work, we investigate the interaction between a plasma jet and the background medium from collisionless to strong collisions. The deflection of the jet is closely linked to the collision states. In the absence of collision or under weak collision conditions, the direction of jet propagation remains largely unchanged. However, in cases of collision or strong collision, the direction of jet propagation undergoes significant deflection. Additionally, we observe a double-shock structure in the optical images and self-emitting X-ray data, where the impact point is pushed inward into the CH background. These observations align with astrophysical simulations \citep{raga2002}. It is important to note that astrophysical jet deflections occur on spatial and temporal scales that are typically 10 to 20 orders of magnitude greater than those achievable in laboratory experiments designed to simulate such phenomena. Therefore, ensuring similarity between astrophysical phenomena and their related laboratory experiments becomes a critical issue. Previous studies \citep{ryutov1999,ryutov2001} have suggested that maintaining key dimensionless numbers at values much larger than unity in both systems, such as the Reynolds number and Péclet number, leads to both systems behaving as ideal compressible hydrodynamic fluids. Moreover, the governing equations remain invariant in the two systems when the following transformation conditions are satisfied: $r_{\text{lab}} = ar_{\text{ast}}$, $\rho_{\text{lab}} = b\rho_{\text{ast}}$, $P_{\text{lab}} = cP_{\text{ast}}$, $t_{\text{lab}} = a\sqrt{b/c} t_{\text{ast}}$, $v_{\text{lab}} = \sqrt{c/b} v_{\text{ast}}$, $B_{\text{lab}} = \sqrt{c} B_{\text{ast}}$, where $a$, $b$, and $c$ are arbitrary positive numbers, also known as free transformation parameters. The subscripts "ast" and "lab" represent the astrophysical and laboratory systems, respectively. 

As presented in Table \ref{scaleparameters}, the astrophysical HH jet showcases negligible viscosity and thermal conductivity due to its vast spatial dimensions, resulting in Reynolds and Péclet numbers that significantly exceed unity. In laboratory experiments, we can compute these crucial dimensionless parameters in the following equations. The Reynolds number can be expressed as $Re = \boldsymbol{v}L/\nu$, where $\boldsymbol{v}$, $L$, and $\nu$ represent velocity, system size, and viscosity, respectively. The viscosity can be determined using the equation $\nu = \left(k_{B}T\right)^{5/2}/[\pi n_{i}m_{i}^{1/2}Z^{4}e^{4}\ln\Lambda]$ \citep{drake2018}, with $k_{B}$, $T$, $n_i$, $m_i$, $Z$, and $\ln\Lambda$ denoting the Boltzmann constant, temperature, ion number density, ion mass, average ionization, and Coulomb logarithm, respectively. The Péclet number is defined as $Pe = vL/\kappa$, where $\kappa = \left(k_{B}T\right)^{5/2}/[m_{e}^{1/2}n_{i}Z(Z+1)e^{4}\ln\Lambda]$ represents the thermal diffusivity \citep{brandenburg2005astrophysical}. In accordance with Table \ref{scaleparameters}, both the Reynolds and Péclet numbers meet the criterion of being significantly greater than unity, thereby confirming the validity of the scaling laws. By selecting transformation coefficients as $a = 3.3 \times 10^{15}$, $b = 2 \times 10^{-17}$, and $c = 3 \times 10^{-18}$, we observe that the scaled parameters of the laser-driven plasma closely resemble those of the HH 110/270 system. Such resemblance between the laser-driven plasma and the astrophysical system reinforces the applicability of our results on the astrophysical scale. Our findings suggest that the complex and chaotic structure observed in the HH 110 jet can be attributed to the Kelvin-Helmholtz instability at the interface. The impact point experiences compression within the dense cloud due to the presence of a backward shock, confirming Raga's deduction that the impact point of HH 270 has shifted deeper into the cloud, westward of the HH 110 axis \citep{raga2002}. These results provide an explanation for the significant $\rm HCO^{+}$ emission observed west of HH 110 \citep{choi2001}, which is associated with the backward shock within the cloud. In instances where the cloud density is sufficiently high, the impact point tends to be in proximity to the contact interface, resulting in additional substantial emissions comparable to the strong-collision scenario witnessed in our experiments.

Our findings present a comprehensive analysis of the interaction process between the plasma jet and background plasma under varying collision states. These results offer valuable insights into understanding the associated shock phenomena observed in the HH deflection system. Moreover, the experimental platform employed in this study can be effectively utilized for investigating other shock configurations, including collisionless shocks involving counter-streaming plasma jets. This controlled laboratory laser-driven platform serves as a crucial complement to astrophysical observations and provides a means for validating the deflection model of HH objects.

\begin{acknowledgments}
The authors thank the staff of the $\rm ShenGuang\ (SG)  - \MakeUppercase{\romannumeral 2}$ laser facility for operating the laser system. This work is supported by the National Key R\&D Program of China, Grants No. 2022YFA1603200, No. 2022YFA1603201 and No. 2022YFA1603204; Z.L. is supported by the China Postdoctoral Science Foundation, No. 2023M730334. This work is also supported National Natural Science Foundation of China, Grant No. 11825502 and 11921006. B.Q. acknowledges support from the National Natural Science Funds for Distinguished Young Scholar, grant No. 11825502. The simulations are carried out on the Tianhe-2 supercomputer at the National Supercomputer Center in Guangzhou.
\end{acknowledgments}

%

\appendix
\renewcommand\thefigure{\Alph{section}\arabic{figure}}    
\setcounter{figure}{0}
\section{Experiment Setup.} \label{app:expSetup}

\begin{figure}[htb]
    \centering
    \includegraphics[width=15cm]{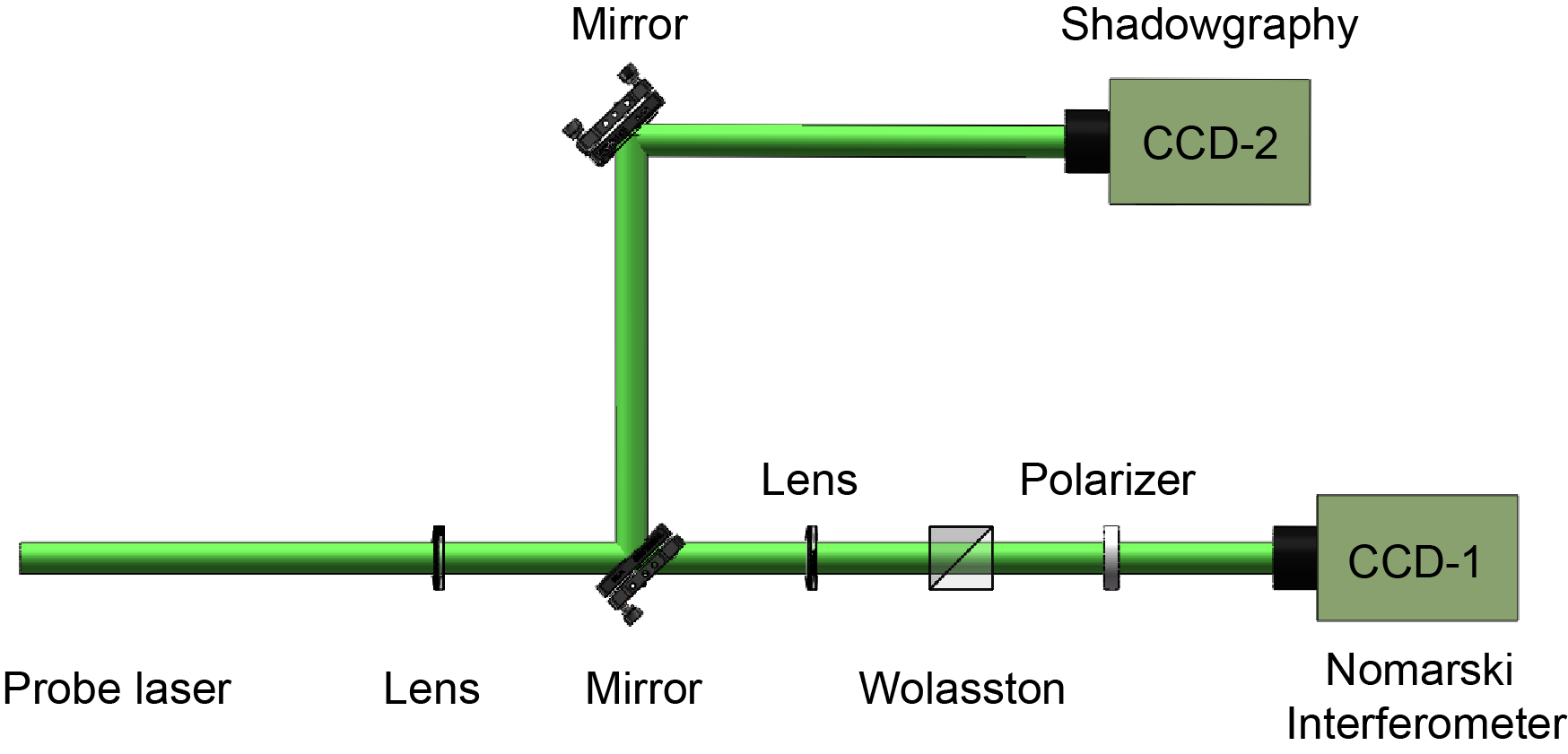} 
    \caption{{\bf Schematic view of the experimental probe laser setup.} The probe laser is split into two light pulses for the Nomarski interferometer and shadowgraphy respectively.}
    \label{lasersetup}
\end{figure}

The experiments were conducted at the ShenGuang-II (SG-II) laser facility in Shanghai, China. The background plasma was generated using either the laser-driven CH plain target or the hemisphere target. To produce the plasma jet, a V-shaped Cu target was utilized, consisting of two Cu foils with a thickness of $\rm 500 ~ \mu m$. Each foil was driven by two laser beams. The driving lasers, with a wavelength of $\rm \lambda_L = 351 ~ nm$, were smoothed using a continuous phase plate (CPP) with an energy of $\rm 250~ J $ over a duration of $\rm 1 ~ ns$. The focal spot diameter of each laser beam was approximately $\rm 600 ~ \mu m$, resulting in an average intensity of $2.5\times 10^{13} ~ W/cm^2$. The main diagnostics employed in the experiment included an optical probe beam at 527 nm, a pinhole X-ray camera, and an electron spectrometer. The detail probe laser setup is shown in Fig.~\ref{lasersetup}, and the probe laser is split into two light pulses for the Nomarski interferometer and shadowgraphy respectively. A pinhole camera was equipped with a $\rm 2 ~ \mu m $-thick aluminum (Al) foil filter, and the electron spectrometer recorded spectra ranging from 20 to 500 keV. The uniform magnetic field in electron spectrometers was set to an intensity of 600 G, generated by two permanent magnets. For more details, please refer to the paper by Liu et al. \citep{liu2017}. 

\section{Radiation-magneto-hydrodynamics simulation setup} \label{app:simSetup}
The radiation–magneto-hydrodynamics code FLASH \citep{fryxell2000} is used to model the laser absorption and the plasma interaction of strong collision case. The target configuration and the laser parameters, such as incident angles, focal size and pulse duration are accordance with the experiments. For the simulation free expansion of CH plasma, a $\rm x\times y \times z =  3 ~ mm \times 3 ~ mm \times 4 ~ mm$ simulation box is set. The adaptive mesh refinement method is applied in the simulations, and the maximum resolution is $\rm 10 ~ \mu m$.  A laser beam is incident on the target surface obliquely. For the simulation of jet-only case, the computational domain is set as $\rm x\times y \times z =  7 ~ mm \times 4 ~ mm \times 4 ~ mm$ with the maximum resolution $\rm 10 ~ \mu m$. For the plasma interaction cases, two-dimensional simulation is performed to achieve higher resolution accuracy. The computational domain of 2D simulation is set as $\rm x\times y = 7 ~ mm \times 4 ~ mm$, and the maximum resolution is about $\rm 4 ~ \mu m$. The self-generated Biermann magnetic field is also include in our simulations \citep{lei2020numerical}. 

\section{Collision Parameter} \label{app:collParameter}
The mean free path ($\lambda_{\alpha \beta}$) for collisions between species $\rm \alpha$ and $\rm \beta$ in the counter-streaming flows can be determined using the following expression \citep{fiuza2020,popovics1997}: $\lambda_{\alpha \beta} = m_{\alpha \beta}^2 v_r^4 \operatorname{ln} \Lambda_{ {\alpha \beta }} / (4\pi Z_{\alpha}^2 Z_{\beta}^2 e^4 n_{\beta}) $. Here, $m_{\alpha \beta}$, $v_r$, $Z_{\alpha ~ \text{or} ~ \beta}$, $\operatorname{ln} \Lambda_{\alpha \beta}$, and $n_{\beta}$ represent the reduced mass, relative velocity, average ionization state, Coulomb logarithm, and mean number density, respectively. The reduced mass can be calculated as $m_{\alpha \beta} = \frac{m_{\alpha}m_{\beta}}{m_{\alpha} + m_{\beta}} = \frac{A_{\alpha}A_{\beta}}{A_{\alpha} + A_{\beta}}$, where $A_{\alpha ~\text{or}~ \beta}$ represents the atomic mass number for different species. Based on simulation and experimental results, the following values can be estimated for Cu and CH plasma: $v_r = 1.2 ~\text{km/s}$, $A_{\alpha} = 64$, $\bar{A}_{\beta} = 6.5$, $\bar{Z}_{\alpha} \approx 10$, $\bar{Z}_{\beta} \approx 3.5$, and $\operatorname{ln} \Lambda_{\alpha \beta} \approx 8$. By substituting these values into the aforementioned formula, the mean free path can be calculated.

\section{PIC simulation setup} \label{app:PICSetup}
In our PIC simulation, the simulation box size is $(L_x, L_y) = 2000 ~ \mathrm{\mu m} \times 2000 ~ \mathrm{\mu m}$, divided into $(N_x, N_y) = 500 \times 500$ grids. We employ 60 macro-particles per species in each grid. The ion skin depth of CH is approximate $d_{i, CH} \approx 70 ~ \mathrm{\mu m}$, and the ratio of $d_{CH,i}$ to the spatial resolution of the simulation box is 17.5. The boundary conditions for particles and fields are set to be open. For the collisionless case, we perform two-dimensional PIC simulations of plasma interaction using the fully electromagnetic and massively parallel PIC code Epoch 4.17 \citep{arber2015}. To better replicate the experimental results, we apply a self-consistent transform method in our simulation. This transformation method involves scaling the plasma states of the experimental data to adapt them to the PIC simulation. The scaling parameters are chosen such that key dimensionless numbers, such as plasma $\beta$ and Mach number, are conserved between the two systems \citep{zhao2022}. To include Coulomb collisions effectively, we ensure that the ratio $L_{\mathrm{system}} / \lambda_{ii}$ is preserved, where $\lambda_{ii}$ represents the ion mean free path. Additionally, the ratio $L_{\mathrm{system}} / d_i$ is kept unchanged, where $d_i = c / \omega_{pi} = c / (4\pi n_i Z^2 e^2 / m_i)^{1/2}$ is the ion skin depth. This ensures that the two-fluid effect remains consistent between the experimental and PIC simulation systems. The self-similar transformation details are as follows:
$$ L_{PIC} = f_L L_{exp},\ m_{i,PIC} = f_m m_{i,exp},\ n_{i,PIC} = f_n n_{i,exp},\ T_{PIC} = f_T T_{exp}  $$
$$t_{PIC} = f_L \sqrt{f_m / f_T} t_{exp},\ v_{PIC} = \sqrt{f_T / f_m} v_{exp},\ B_{PIC} = \sqrt{f_n f_T} B_{exp} $$
Here, $L, m, n, T$ represent the size, ion mass number, number density, and temperature, respectively, while $t, v, B$ represent time, velocity, and magnetic field strength. We choose $f_n = 1$ to maintain the number density unchanged between the two systems, and $f_m = 0.05$ to reduce the ion mass number in the PIC simulation. To preserve $L_{\mathrm{system}} / d_i$, we determine that $f_r = \sqrt{f_m / f_n}$. In our simulation, we set $f_T = 5$. With these transformation parameters, the PIC simulation can be accelerated while maintaining the essential characteristics of the system.

\section{Integration Method for Self-emitting X-rays} \label{app:Xrays}
The presence of a $\rm 2 ~ \mu m $-thick aluminum (Al) foil filter in the pinhole camera restricts the recording of X-rays to energies above approximately $\sim 400 ~ eV$. According to the Bethe-Heitler formula \citep{brown1971}, the differential cross-section $\rm Q_{\varepsilon}(E)$ (where $\rm E$ represents the electron energy) for photon energy $\varepsilon$ can be expressed as $\rm Q_{\varepsilon}(E)$ (where $\rm E$ is the electron energy) of photon energy $\varepsilon$ is $Q_{\varepsilon}(E)=\frac{8}{3} \frac{r_{0}^{2}}{137} \frac{m c^{2}}{\varepsilon E} \log \frac{1+\sqrt{1-\varepsilon / E}}{1-\sqrt{1-\varepsilon / E}}$. Here, $m$ and $r_0$ denote the rest mass and classical radius of the electron, respectively, while $c$ represents the speed of light. Consequently, photons with an energy of 400 eV are predominantly emitted by electrons with temperatures exceeding 800 eV. Therefore, the integration is performed within the region where the temperature exceeds 800 eV. The power per unit volume of self-emitting X-rays scales as $T^{1/2} n_e n_i Z^2 \ (erg ~ s^{-1} ~ cm^{-3})$. Consequently, the data corresponding to different time points is integrated to obtain the total emitting images in the simulations.

\bibliography{sample631}{}
\bibliographystyle{aasjournal}



\end{document}